\documentclass[review]{elsarticle}

\usepackage{lineno,hyperref}

\usepackage[margin=1in]{geometry}

\pdfoutput=1

\usepackage{graphicx}
\usepackage{bm}
\usepackage{dcolumn}
\usepackage{setspace}
\usepackage{amsmath}
\usepackage{fancyhdr}

\usepackage{caption}
\usepackage{subcaption}

\makeatletter
\renewcommand{\fnum@figure}{Fig. \thefigure}
\makeatother

\journal{Computational Condensed Matter}

\begin{document}

\begin{frontmatter}

\title{Effect of orientation and mode of loading on deformation behaviour of Cu nanowires}


\author[mymainaddress]{P. Rohith}

\author[mymainaddress]{G. Sainath\corref{mycorrespondingauthor}}
\cortext[mycorrespondingauthor]{Corresponding author}
\ead{sg@igcar.gov.in}

\author[mymainaddress]{B.K. Choudhary}

\address[mymainaddress]{Materials Development and Technology Division, Metallurgy and Materials Group, Indira Gandhi Centre for Atomic Research, HBNI, Kalpakkam, Tamilnadu-603102, India}

\begin{abstract}
Molecular dynamics simulations have been performed to understand the variations in deformation mechanisms of 
Cu nanowires as a function of orientation and loading mode (tension or compression). Cu nanowires of different 
crystallographic orientations distributed uniformly on the standard stereographic triangle have been considered 
under tensile and compressive loading. The simulation results indicate that under compressive loading, the 
orientations close to $<$100$>$ corner deform by twinning mechanism, while the remaining orientations deform 
by dislocation slip. On the other hand, all the nanowires deform by twinning mechanism under tensile loading.
Further, the orientations close to $<$110$>$ and $<$111$>$ corner exhibit tension-compression asymmetry in 
deformation mechanisms. In addition to deformation mechanisms, Cu nanowires also display tension-compression 
asymmetry in yield stress. The orientations close to $<$001$>$ corner exhibits higher yield stress in tension 
than in compression, while the opposite behaviour (higher yield stress in compression than in tension) has been 
observed in orientations close to $<$110$>$ and $<$111$>$ corners. For the specific orientation of $<$102$>$, 
the yield stress asymmetry has not been observed. The tension-compression asymmetry in deformation mechanisms 
has been explained based on the parameter $\alpha_M$, defined as the ratio of Schmid factors for leading and 
trailing partial dislocations. Similarly, the asymmetry in yield stress values has been attributed to the 
different Schmid factor values for leading partial dislocations under tensile and compressive loading.
\end{abstract}

\begin{keyword}
Molecular Dynamics simulations, Cu nanowire, Orientation, Dislocations and Twinning
\end{keyword}

\end{frontmatter}


\section{Introduction}
In recent years, the metallic nanowires have attracted a significant interest for research due to their superior electrical, 
optical, thermal and mechanical properties. In particular the Cu nanowires/nanopillars, nanobelts, nanosprings and nanofilms 
have emerged as the next-generation materials in nano/micro electromechanical systems (NEMS/MEMS) due to their excellent 
performance with conductivity, transmittance, mechanical flexibility, low cost, easy and inexpensive synthesis \cite{Lieber,NASA}. 
In view of this, understanding the deformation behaviour of Cu nanowires becomes essential for their effective practical 
applications. The knowledge of deformation mechanisms also becomes important for fine-tuning the physical properties of 
nanowires such as reorientation, shape-memory and pseudo-elasticity \cite{Park-PRL}, which have practical implications in 
the design of novel and flexible NEMS/MEMS devices. 

In view of the small size of nanowires, performing the accurate mechanical testing of nanowires is still challenging due to 
the difficulties in sample preparation, clamping and aligning the nanowire axial direction with loading direction \cite{Review-APR}. 
This complexities in experimental techniques preclude the conventional methods and lend towards theoretical/computational 
tools. With the rapid progress of computational capability and the availability of reliable inter-atomic potentials, molecular 
dynamics (MD) simulations have become a major tool to probe the mechanical properties of nanoscale materials. In addition to 
the usual mechanical properties, MD simulations provide the real-time deformation process of nanowire at the atomic scale. 
In the present study, MD simulations have been used to understand the deformation mechanisms in Cu nanowires.

In the past, many experimental/atomistic simulations have been performed on FCC nanowires/nanopillars such as Au 
\cite{Park-JMPS,Zheng-Nature,Lee-Nature}, Ag \cite{Park-Acta-Ag}, Cu \cite{Park-JMPS,Cao-Acta-Cu}, Pt \cite{Pt-NL}, 
Al \cite{Al-IJP} and Ni \cite{Park-JMPS,Horstemeyer-IJP}. All these studies have shown that the important mechanisms 
of plastic deformation in FCC nanowires are slip through perfect and partial dislocations and deformation twinning. 
The competition between slip and twinning mechanisms depends mainly on the crystallographic orientation, size, shape 
and loading mode (tension/compression), temperature and strain rate \cite{Cai-Rev,Mao-EML,Xie-Srate}. However, among 
all these factors, crystallographic orientation has the strongest influence followed by mode of loading 
\cite{Horstemeyer-IJP,Cai-Rev}. For example, Zheng et al. \cite{Zheng-Nature} studied the orientation dependent deformation 
behaviour of Au nanocrystals using 
in-situ high resolution transmission electron microscope (HRTEM) combined with MD simulations. It has been shown that under 
tensile loading of Au nanocrystal, the deformation by slip is favoured in $<$100$>$ orientation, while twinning is observed 
in $<$110$>$ orientation. This study further confirmed that, despite large differences in strain rates, the results obtained 
using MD simulations are in good agreement with experimental results \cite{Zheng-Nature}. Similar to $<$110$>$ orientation, 
the deformation by partial slip/twinning is preferred under the tensile loading of $<$111$>$ oriented FCC nanowires 
\cite{Cao-Acta-Cu}. In order to understand the influence of mode of loading (tension/compression), Lee et al. \cite{Lee-Nature} 
carried out the experimental and atomistic simulation study on the tensile and compressive deformation of $<$110$>$ Au 
nanopillars. Under tensile loading, the deformation by twinning was observed, while dislocation slip has occurred under 
the compressive loading of $<$110$>$ Au nanopillar \cite{Lee-Nature}. Contrary to $<$110$>$ orientation, the $<$100$>$ 
oriented FCC nanowires deform by slip under tensile loading, while twinning is observed under compressive loading 
\cite{Park-JMPS,Zheng-Nature,Cai-Rev}. These results strongly suggest that the deformation behaviour of FCC nanowires 
depends on the orientation and mode of loading. However, in the literature, the deformation mechanisms and tension-compression 
asymmetry have been investigated mainly in high symmetry $<$100$>$, $<$110$>$ and $<$111$>$ crystallographic orientations, 
which constitutes the corners of a standard stereographic triangle \cite{Cai-Rev}. In view of this, understanding the 
deformation mechanisms over a wide range of orientations in FCC nanowires becomes important. Moreover, the variations of 
yield strength and strength asymmetry with respect orientation have not been studied systematically.

In the present study, MD simulations have been performed on Cu nanowires of different crystallographic orientations 
distributed uniformly on the standard stereographic triangle. The orientations falling on the corners, boundaries and in the 
interior of the standard stereographic triangle have been considered. In addition to above, the influence 
of loading mode (tension/compression) has been examined for all the orientations. The yielding behaviour, yield strength and 
deformation mechanisms of all the nanowires under tensile and compressive loadings have been presented. Finally an attempt has 
been made to understand the orientation dependent deformation mechanisms in terms of the ratio of leading partial Schmid 
factor to trailing partial Schmid factor.  

\section{MD Simulation details}

Molecular dynamics (MD) simulations have been performed using Large scale Atomic/Molecular Massively Parallel Simulator 
(LAMMPS) package \cite{Plimpton-1995} employing an embedded atom method (EAM) potential for FCC Cu given by Mishin and 
co-workers \cite{Mishin-2001}. This potential has been chosen for being able to reproduce stacking fault and twinning 
fault energies for Cu \cite{Liang-PRB}, which are key variables for predicting the dislocation nucleation and deformation 
mechanisms in nanowires. 

\begin{figure}[h]
\centering
\includegraphics[width=6cm]{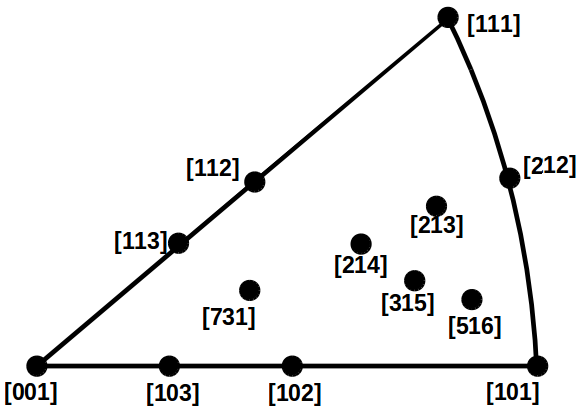}
\caption {Standard stereographic triangle showing the different orientations considered in this study. Cu nanowires 
with these orientations have been subjected to tensile and compressive loading.}
\label{Fig01}
\end{figure}

In order to examine the influence of orientation on the deformation behaviour, MD simulations have been performed on Cu 
nanowires of different crystallographic orientations distributed uniformly over the standard stereographic triangle 
(Fig. \ref{Fig01}). The orientations falling on the corners ($<$100$>$, $<$110$>$ and $<$111$>$), boundaries ($<$115$>$, 
$<$113$>$, $<$112$>$, $<$212$>$, $<$102$>$ and $<$103$>$) and interior ($<$213$>$, $<$214$>$, $<$315$>$ and $<$516$>$) 
of the standard stereographic triangle have been considered. The side surfaces of all the nanowires 
with different orientations have been presented in Table \ref{Side-surfaces}. All the nanowires have a square cross-section 
width (d) of 10 nm and length of 20 nm, providing an aspect ratio of 2:1. Periodic boundary conditions have been chosen 
along the nanowire length direction, while the other two directions were kept free in order to mimic an infinitely long 
nanowire. Following initial construction of the nanowire, energy minimization was performed by conjugate gradient method 
to obtain a relaxed structure. The relaxed nanowire is thermally equilibrated to a required temperature of 10 K in canonical 
ensemble (constant NVT) with a Nose-Hoover thermostat. Velocity verlet algorithm has been used to integrate the equations of 
motion with a time step of 2 fs. 

Upon completion of the equilibration process, the nanowires were deformed under tensile and compressive loading at a 
constant engineering strain rate of $1 \times 10^8$ s$^{-1}$  along the nanowire length direction. The strain rate
considered in MD simulations is significantly higher than the experimental strain rates, which is due to the inherent 
time scale limitations from molecular dynamics. The strain ($\varepsilon$) has been calculated as $(l-l_0)/l_0$, where 
$l$ is instantaneous length and $l_0$ is the initial length of the nanowire. The stress has been obtained using the 
Virial expression \cite{Virial}, which is equivalent to a Cauchy's stress in an average sense. Finally, AtomEye package 
\cite{AtomEye} and OVITO \cite{OVITO} have been used for the visualisation of atomic snapshots with common neighbour 
analysis (CNA) \cite{CNA1,CNA2}.

\begin{table}[h]
\caption{The nanowire orientations along with their side surfaces.}
\vspace{0.2cm}
\label{Side-surfaces}
\centering
\scalebox{0.75}{
\begin{tabular}{| l | l | l | l | l | l | l | l | l | l | l | l| l |} \hline
\rule{0pt}{3ex}
Orientation & $<$001$>$ & $<$101$>$ & $<$102$>$ & $<$103$>$ & $<$111$>$ & $<$112$>$ & $<$113$>$ & $<$212$>$ & $<$213$>$ 
& $<$214$>$ & $<$315$>$ & $<$516$>$ \\ [3pt]
\hline
\rule{0pt}{3ex}
Side surfaces & \{100\} & \{100\} & \{100\} & \{100\} & \{110\} & \{110\} & \{110\} & \{110\} & \{111\} & \{201\} & \{130\} & 
\{111\} \\ [3pt]
\rule{0pt}{3ex}
 & \{100\} & \{110\} & \{120\} & \{130\} & \{112\} & \{111\} & \{233\} & \{114\} & \{145\} & \{1,2,10\} & \{123\} & \{4,7,11\} \\ 
 [3pt]
\hline 
\end{tabular} }
\end{table}

\section{Results}

\subsection{Deformation behaviour under compression}

The compressive loading has been performed on Cu nanowires of different crystallographic orientations up to a strain 
value of 0.4. The evolution of atomic configurations at various stages of deformation were analysed for all the 
orientations using the common neighbour analysis. Based on this analysis, the deformation mechanisms under the compressive 
loading of Cu nanowires have been classified into twinning and dislocation slip as described in the following:

\subsubsection{Deformation through twinning}

\begin{figure}[h]
\centering
\includegraphics[width=9cm]{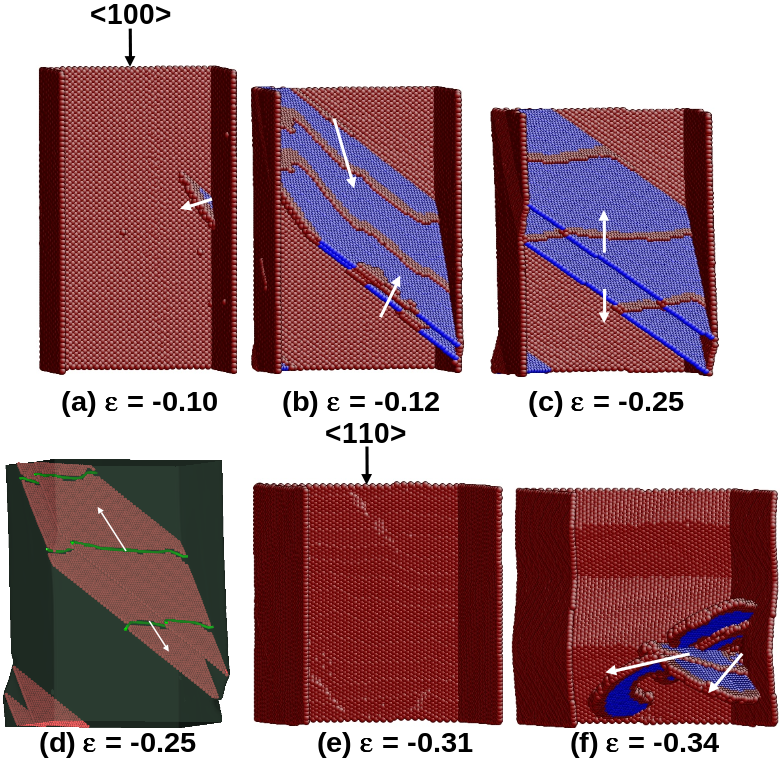}
\caption {Atomic snapshots displaying (a) nucleation of a Shockley partial, (b) glide of twinning partial and 
micro-twin, (c-d) twin growth (e) reoriented nanowire, and (f) the slip of extended dislocation in the reoriented nanowire 
during the compressive deformation of $<$100$>$ Cu nanowires. In Figure (d), the partial dislocations are shown as green lines 
as obtained in OVITO, while the remaining figures are obtained using AtomEye. The atoms are coloured according to the common 
neighbour analysis (CNA). The blue colour atoms represents the FCC and the red colour atoms indicate the surfaces and dislocation 
core.}
\label{100-compression}
\end{figure}

Cu nanowires with $<$100$>$, $<$103$>$ and $<$113$>$ orientations (Fig. \ref{Fig01}) deforms by twinning mechanism 
under compressive loading. As a representative of twinning mechanism, Fig. \ref{100-compression} shows the deformation 
behaviour of $<$100$>$ Cu nanowire at different strain levels. It can be seen that the nanowire yields through the nucleation 
of a single 1/6$<$112$>$ Shockley partial dislocation on \{111\} plane from the corner with a stacking fault in its wake
(Fig. \ref{100-compression}a). 
On the contrary, nucleation of multiple Shockley partial dislocations has been observed during the compressive deformation of 
$<$103$>$ and $<$113$>$ Cu nanowires. Following yielding, another Shockley partial nucleates on the adjacent plane and 
creates micro-twin in the nanowire (Fig. \ref{100-compression}b). This shows that the presence of stacking faults are 
must for formation of twins in the nanowires \cite{Kardani1,Kardani2}. The stacking faults needed to form the twins were 
created by the glide of leading partials nucleated from the nanowire surface. With increasing strain, the continuous nucleation and 
glide of twinning partials along the twin boundaries leads to twin growth (Fig. \ref{100-compression}c). It can be seen 
that the twinning partials glide in mutually opposite directions on two twin boundaries and as a result the twin boundaries 
move away from each other leading to the twin growth (Fig. \ref{100-compression}c). The continuous propagation of twin 
boundaries along the nanowire length progressively reorients the twinned region. Due to periodic boundary conditions along 
the nanowire length, the twin boundaries across the length meet each other and annihilate leaving a defect free reoriented 
nanowire (Fig. \ref{100-compression}d). Thus, the deformation by twinning on a single twin system transforms the orientation 
of nanowire from $<$100$>$/\{100\} to $<$110$>$/\{111\}. After the reorientation, the nanowire again undergoes an elastic 
deformation followed by yielding through the nucleation of a new 1/6$<$112$>$ Shockley partial on a different crystallographic 
plane. Following yielding, the deformation in the reoriented nanowire proceeds by the slip of extended dislocations (Fig. 
\ref{100-compression}e). Similar to $<$100$>$ nanowire, deformation by partial slip and twinning has been observed under 
the compressive loading of $<$103$>$ and $<$113$>$ Cu nanowires. However, due to twin-twin interactions, the reorientation 
has not been observed in $<$103$>$ and $<$113$>$ nanowires.  

\subsubsection{Deformation through slip}

\begin{figure}[h]
\centering
\includegraphics[width=12cm]{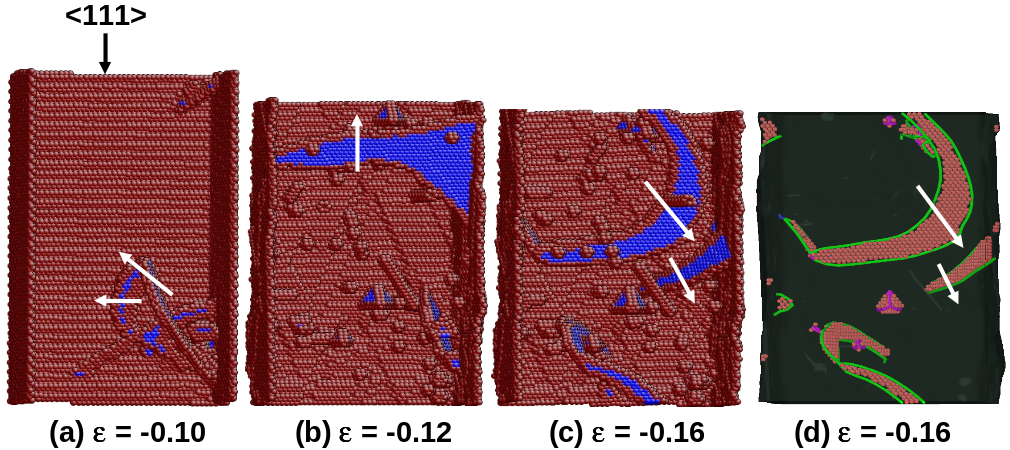}
\caption {Atomic snapshots displaying the compressive deformation of $<$111$>$ Cu nanowires as a function of 
strain. The deformation by the slip of extended dislocations along with stacking fault tetrahedron and many point defects 
can be seen in (b), (c) and (d). Figure (d) is the OVITO output of Figure (c), where partial dislocations are shown as green 
lines, the stair-rod dislocations enclosing the stacking fault tetrahedron are shown in magenta lines. The atoms are coloured 
according to the common neighbour analysis (CNA).}
\label{111-compression}
\end{figure}

\begin{figure}[h]
\centering
\includegraphics[width=10cm]{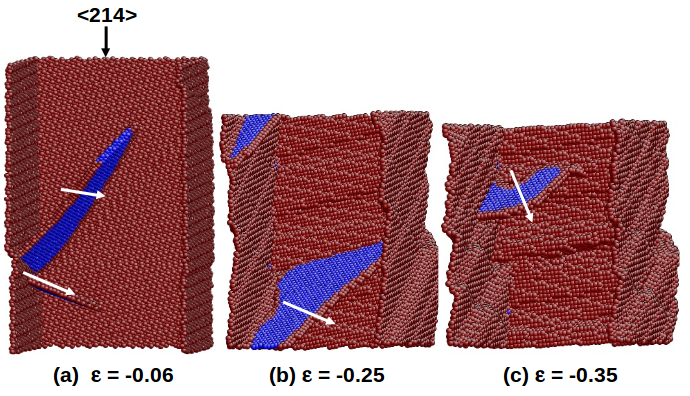}
\caption {Atomic snapshots displaying the compressive deformation of $<$214$>$ Cu nanowires as a function of 
strain. The deformation proceeds through the glide of extended dislocations. The atoms are coloured according to the 
common neighbour analysis (CNA).}
\label{214-compression}
\end{figure}

The Cu nanowires with $<$101$>$, $<$111$>$, $<$102$>$, $<$212$>$, $<$213$>$, $<$214$>$, $<$315$>$ and $<$516$>$ orientations 
deform by full dislocation slip mechanism. As an example, the deformation behaviour in $<$111$>$ (representing the orientations 
along the boundary of a triangle in Fig. \ref{Fig01}) and $<$214$>$ (representing the orientations inside the triangle in 
Fig. \ref{Fig01}) Cu nanowires has been presented in Fig. \ref{111-compression} and \ref{214-compression}, respectively. 
It can be seen that the yielding in both the nanowires occurs through the nucleation of a leading partial immediately followed 
by trailing partial, thus constituting an extended dislocation (Fig. \ref{111-compression}a and \ref{214-compression}a). 
However in $<$111$>$ nanowire, the activation of many slip planes has been observed, while in $<$214$>$ nanowire, the activation 
of only two slip systems has been observed. Following yielding, the nucleated dislocations glide from one corner 
of the nanowire to the opposite and gets annihilated by leaving a slip steps on the surface of the nanowire (Figs. 
\ref{111-compression}b and \ref{214-compression}b). The activation of different slip systems in $<$111$>$ nanowire felicitates 
dislocation-dislocation interactions, which results in the formation of many point defects and also the stacking fault 
tetrahedron as shown in Fig. \ref{111-compression}b-c. Due to the activation of limited slip system in $<$214$>$ nanowire, 
the formation of point defects has not been observed in the strain range examined (Fig. \ref{214-compression}b-c). Generally, 
the stacking fault tetrahedron (SFT) in low stacking fault energy materials can form through vacancy condensation, or by 
an extension of Frank partial dislocation loop or through dislocation-dislocation interactions. However, the SFT formation 
through vacancy condesation requires longer time scales, which are difficult to access using MD simulations. In the present 
study, the SFTs have been formed through dislocation-dislocation interactions, similar to that observed by Wang et al. \cite{SFT-formation}.

\subsection{Deformation behaviour under tension}

In order to investigate the influence of loading mode on the deformation behaviour, MD simulations have been performed on 
the tensile loading of Cu nanowires with different orientations shown in Fig. \ref{Fig01}. It has been observed that 
under tensile loading, the deformation in all the Cu nanowires is dominated by twinning mechanism irrespective of the 
orientations. Full dislocation slip has not been observed in any of the orientations examined. Further, it has been 
observed that the twinning in certain orientations leads to complete reorientation of the nanowires, while in remaining 
orientations, it leads to neck formation and early failure without any reorientation. Therefore, the tensile deformation 
of Cu nanowires has been classified into (i) twinning leading to full reorientation of the nanowire and (ii) twinning 
without full reorientation.

\subsubsection{Twinning leading to full reorientation of the nanowire}

\begin{figure}[h]
\centering
\includegraphics[width=10cm]{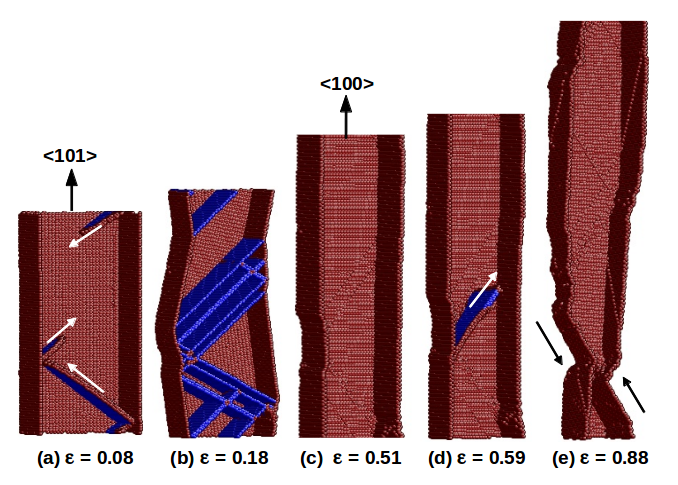}
\caption {The atomic snapshots displaying the tensile deformation of $<$110$>$ Cu nanowires as a function of strain.
Deformation proceeds through twinning followed by reorientation to $<$100$>$ axial direction. }
\label{110-tension}
\end{figure}

\begin{figure}[h]
\centering
\includegraphics[width=10cm]{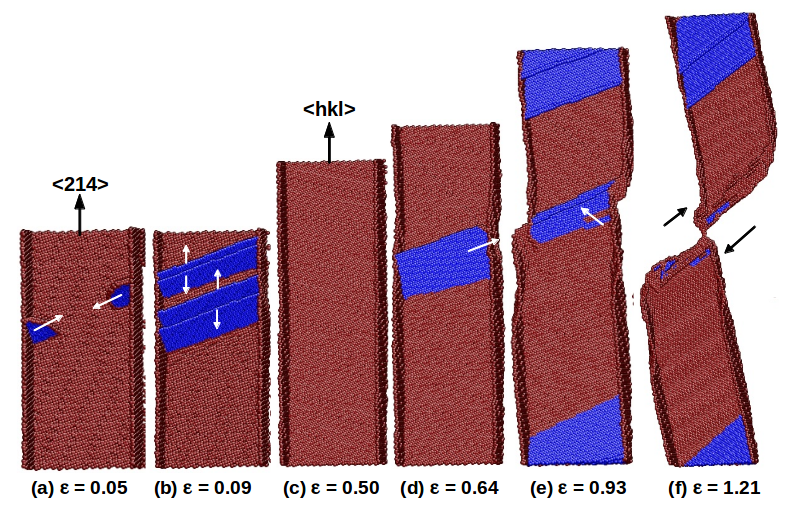}
\caption {The atomic snapshots during the tensile deformation of $<$214$>$ Cu nanowires as a function of strain.
Deformation proceeds through twinning mechanism followed by reorientation to high index $<$hkl$>$ direction. The final 
failure occurs by sliding along the twin boundaries.}
\label{214-tension}
\end{figure}

Twinning under tensile loading of $<$101$>$, $<$103$>$, $<$212$>$ and $<$214$>$ orientations leads to complete reorientation 
of the nanowires. As an example, the deformation by twinning followed by reorientation in $<$101$>$ and $<$214$>$ nanowires has 
been presented in Figs. \ref{110-tension} and \ref{214-tension}, respectively. It can be seen that the yielding in both the 
orientations occurs by the nucleation of Shockley partial dislocations (Figs. \ref{110-tension}a and \ref{214-tension}a).
However in $<$101$>$ nanowire, the activation of two different slip planes has been observed (Fig. \ref{110-tension}a), while 
in $<$214$>$ nanowire, the activation of only one slip system can be seen in Fig. \ref{214-tension}a. Following the nucleation 
of Shockley partials during yielding, the subsequent nucleation of twinning partial dislocations leads to the formation of twins 
on the corresponding \{111\} planes (Figs. \ref{110-tension}b and \ref{214-tension}b). With increasing strain, the twin grows 
along the nanowire axis leading to complete reorientation of nanowires (Figs. \ref{110-tension}c and \ref{214-tension}c). Due 
to twinning, the $<$101$>$ nanowire reorients to $<$100$>$ nanowire, while $<$214$>$ nanowire reorients to a high indexed $<$hkl$>$ 
direction. Generally, the reorientation is not observed when the slip is activated on more than one twin system, due to twin-twin 
interactions. However, the $<$101$>$ nanowire undergoes reorientation despite the activation of multiple twin systems (Fig. 
\ref{110-tension}a-c). Following the reorientation, the nanowires undergo a second elastic deformation followed by yielding through 
the nucleation of Shockley partial dislocations (Figs. \ref{110-tension}d and \ref{214-tension}d). Following the second yielding 
in $<$101$>$ nanowire, twin formation and the interaction of twin boundaries with Shockley partial dislocations results in neck 
formation and final failure as shown in Fig. \ref{110-tension}e. On the contrary, the sliding along the twin boundaries (Fig. 
\ref{214-tension}e) leading to shear failure (Fig. \ref{214-tension}f) accompanied with large strain to failure has been observed 
in $<$214$>$ nanowire.

\subsubsection{Twinning without full reorientation}

\begin{figure}[h]
\centering
\includegraphics[width=10cm]{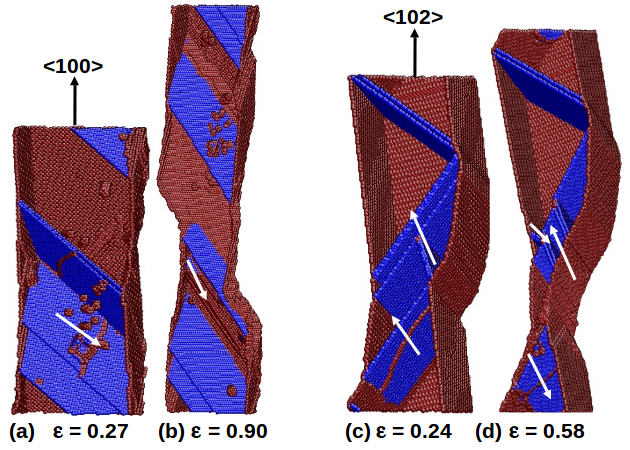}
\caption {Atomic snapshots presenting the deformation and failure behaviour of $<$100$>$ and $<$102$>$ Cu nanowires 
under tensile loading. The twin-twin interactions and the twin boundary sliding disrupting the reorientation process can be 
seen in both the orientations.}
\label{100-102-tension}
\end{figure}

Unlike the $<$101$>$, $<$103$>$, $<$212$>$ and $<$214$>$ orientations, the twinning in $<$100$>$, $<$102$>$, $<$111$>$, $<$112$>$ 
and $<$113$>$ orientations doesn't lead to reorientation. Generally, when the slip is activated on more than one twin system, 
there is an increased probability of twin-twin interactions, which disrupts the twin growth process and reorientation 
\cite{Sainath-CMS15}. Similarly, when the sliding 
along the twin boundaries is preferred over twin boundary migration (which is necessary for twin growth), the reorientation is not 
observed. As an example, the twin-twin interactions and the twin boundary sliding disrupting the reorientation process in $<$100$>$ 
and $<$102$>$ Cu nanowires has been demonstrated in Fig. \ref{100-102-tension}. In $<$100$>$ nanowire, the twin boundary sliding 
dominates over twin growth (Fig. \ref{100-102-tension}a-b), thus making the nanowire to show shear failure without reorientation. 
On the other hand in $<$102$>$ nanowire, the slip-twin interactions along with the sliding of twin boundaries were responsible for 
the absence of reorientation (Fig. \ref{100-102-tension}c-d). 

\subsection{Stress-strain behaviour}

\begin{figure}[h]
\centering
\graphicspath{ {Figures/Chapter5/} }
\begin{subfigure}[b]{0.49\textwidth}
\includegraphics[width=\textwidth]{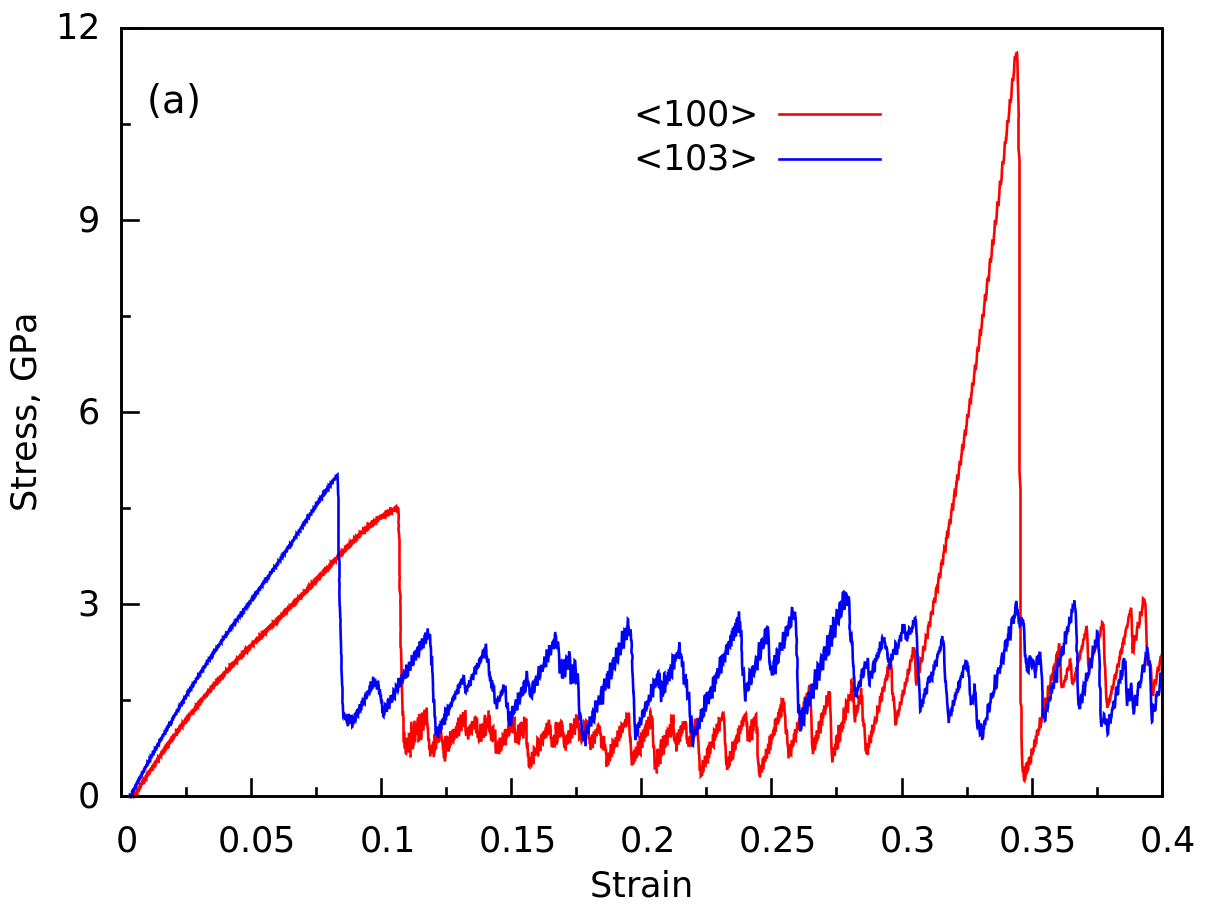}
\end{subfigure}
\begin{subfigure}[b]{0.49\textwidth}
\includegraphics[width=\textwidth]{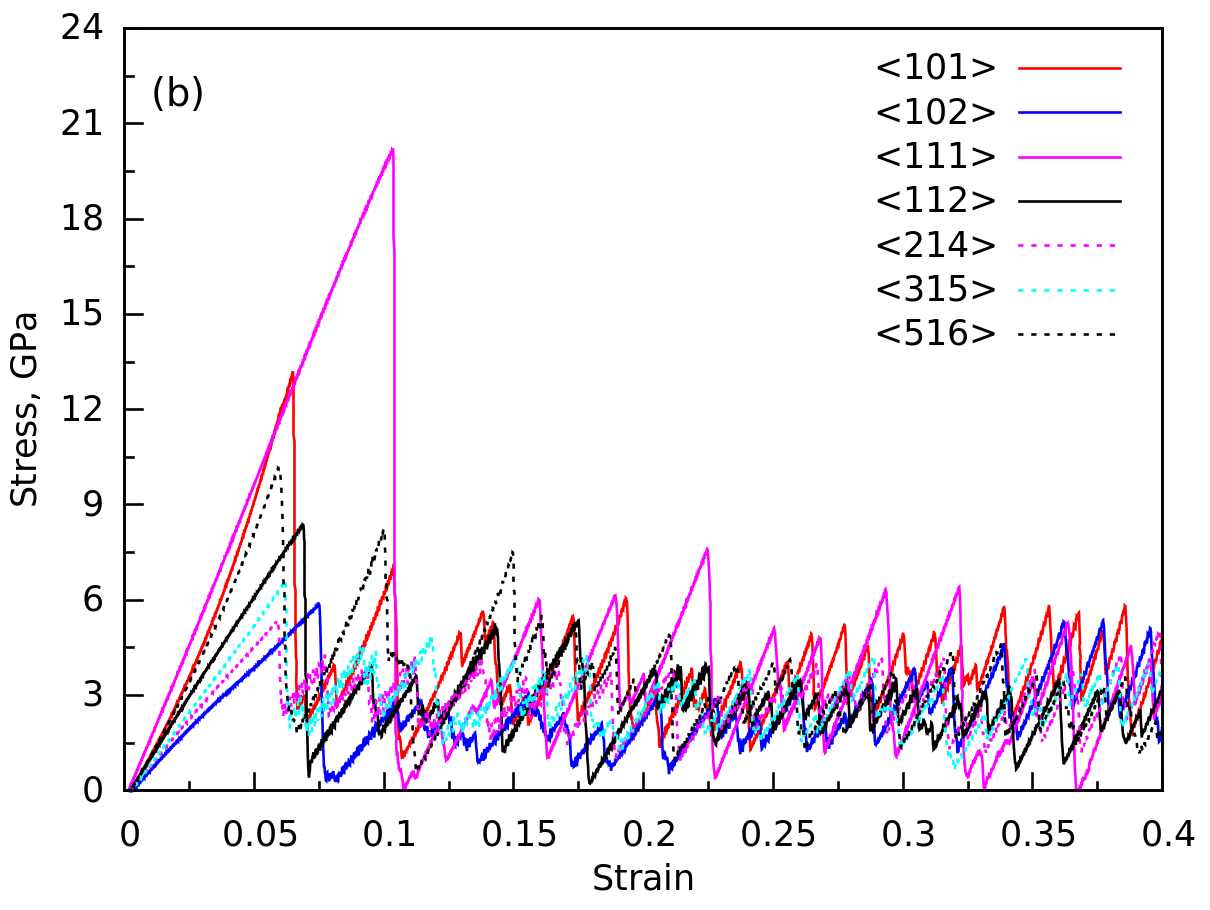}
\end{subfigure}
 \caption {The stress-strain behaviour of Cu nanowires deforming by (a) twinning and (b) full dislocation slip under 
 compressive loading of different orientations at 10 K.} 
 \label{S-S-Compression}
 \end{figure}

Figure \ref{S-S-Compression} shows the stress-strain behaviour of Cu nanowires of different orientations under compressive 
loading. It can be seen that initially all the nanowires undergo elastic deformation up to a peak value followed by an abrupt 
drop in flow stress. Following the yield drop, the flow stress fluctuates about a constant mean value. In nanowires deforming 
by twinning mechanism, marginal fluctuations have been observed about a mean value of 1.3 GPa for $<$100$>$ nanowire and 2.2 
GPa for $<$103$>$ nanowire (Fig. \ref{S-S-Compression}a). Following this fluctuations about a mean value, $<$100$>$ 
nanowire shows second large peak in the stress-strain curve, which is due to the twinning induced reorientation (Fig. 
\ref{S-S-Compression}a). On the other hand in nanowires deforming by full dislocation slip, the flow stress shows large 
fluctuations about a constant mean value of 3 GPa irrespective of the nanowire orientation (Fig. \ref{S-S-Compression}b). 

\begin{figure}[h]
\centering
\graphicspath{ {Figures/Chapter5/} }
\begin{subfigure}[b]{0.49\textwidth}
\includegraphics[width=\textwidth]{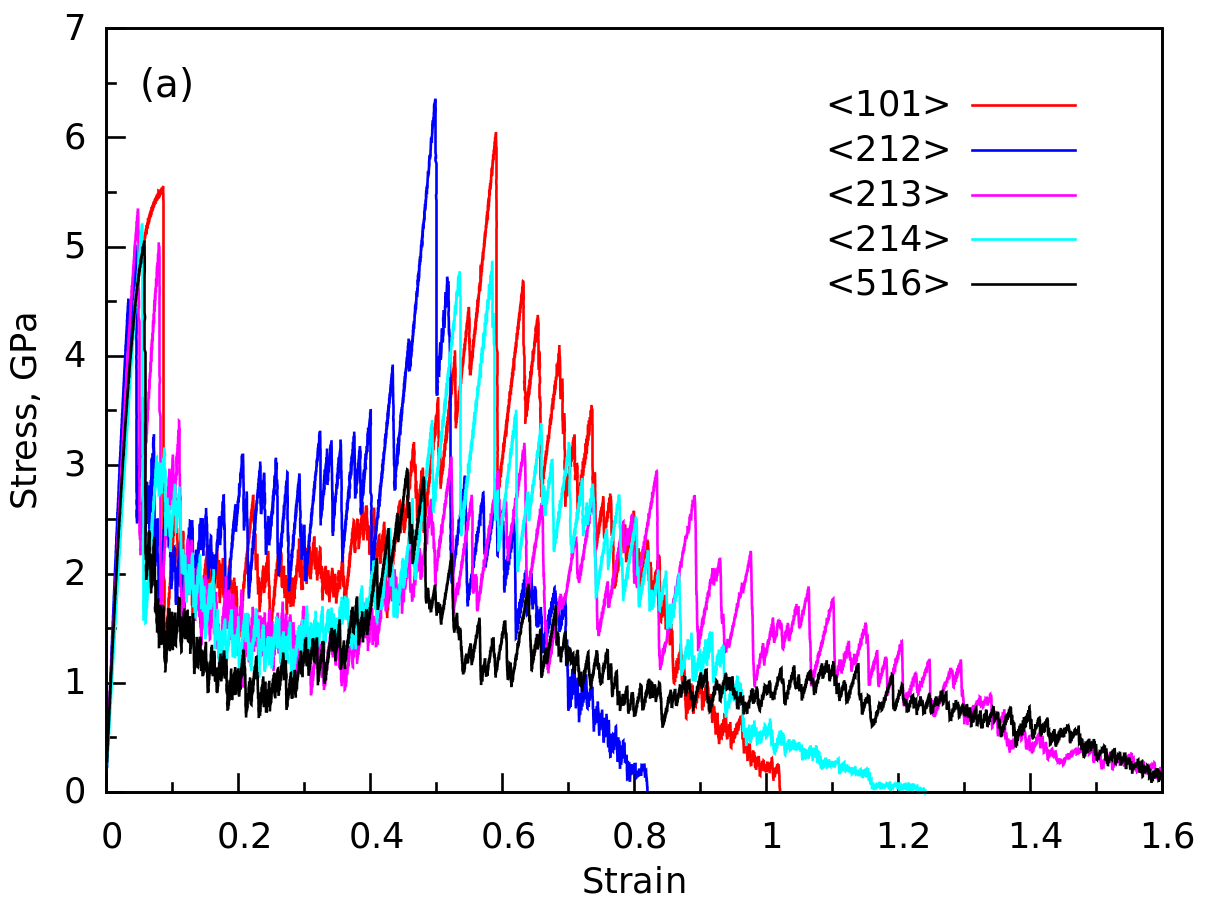}
\end{subfigure}
\begin{subfigure}[b]{0.49\textwidth}
\includegraphics[width=\textwidth]{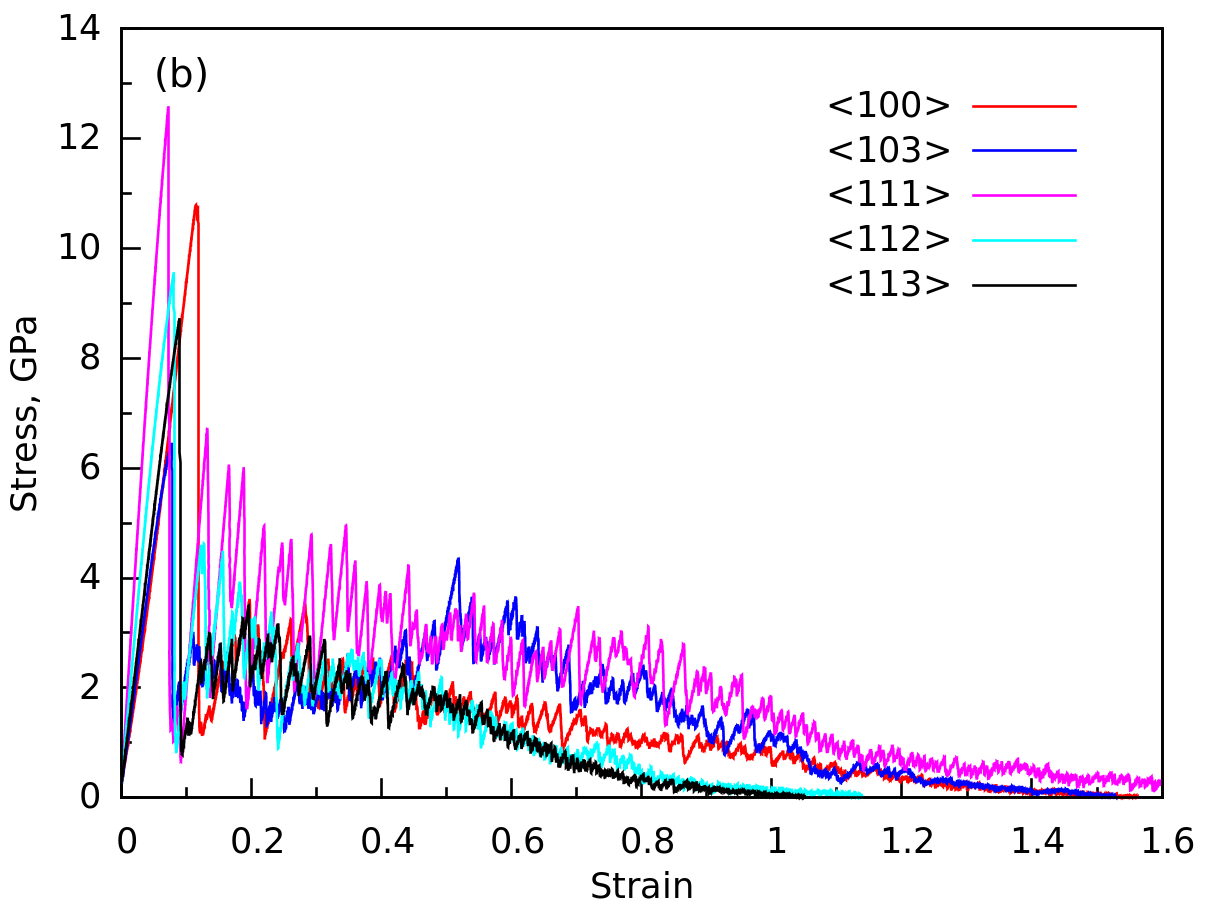}
\end{subfigure}
 \caption {The stress-strain behaviour of Cu nanowires deforming by twinning (a) with reorientation and (b) without 
reorientation under tensile loading at 10 K.} 
 \label{S-S-tension}
 \end{figure}

Figure \ref{S-S-tension} shows the stress-strain behaviour of Cu nanowires with different orientations under tensile loading. 
It can be seen that the stress-strain behaviour of nanowires undergoing a twinning induced reorientation 
(Fig. \ref{S-S-tension}a) is different from that of the nanowires not showing the reorientation (Fig. \ref{S-S-tension}b).
During plastic deformation of nanowires not showing reorientation, the flow stress decreases continuously till failure 
(Fig. \ref{S-S-tension}b), while the flow stress in nanowires undergoing reorientation exhibits a ``U'' shape behaviour 
(Fig. \ref{S-S-tension}a), where the second peak corresponds to the elastic deformation of the reoriented nanowires. Following 
the second peak, the flow stress decreases continuously till failure, similar to the nanowires not showing reorientation. Despite 
similar deformation mechanism by twinning, large variations in strain failure can be seen in Cu nanowires of different orientations 
under tensile loading. 

\begin{figure}[h]
\centering
\graphicspath{ {Figures/Chapter5/} }
\begin{subfigure}[b]{0.40\textwidth}
\includegraphics[width=\textwidth]{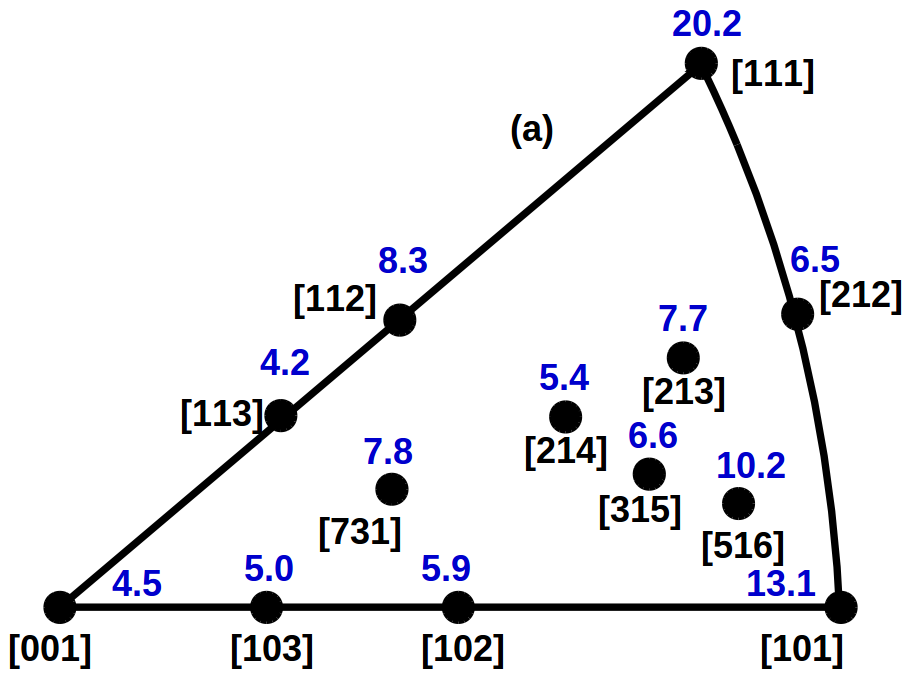}
\end{subfigure}
\begin{subfigure}[b]{0.40\textwidth}
\includegraphics[width=\textwidth]{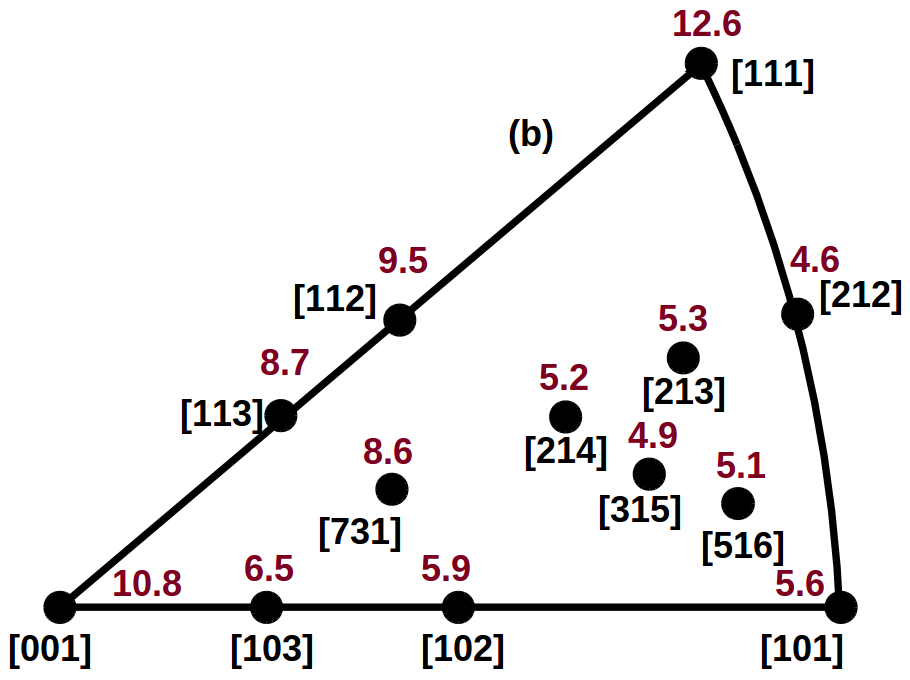}
\end{subfigure}
 \caption {The variation of yield stress with respect to orientation presented on the standard stereographic triangle 
 under (a) compressive loading, and (b) tensile loading of Cu nanowires. The yield stress values are in GPa.} 
 \label{Yield-stress}
 \end{figure}

Following the yielding, the peak value in the stress-strain curve has been taken as the yield stress for Cu nanowires of 
different orientations and presented on the standard stereographic triangle under compressive and tensile loading as shown 
in Fig. \ref{Yield-stress}. Under compressive loading, it can be seen that the Cu nanowires with orientation close to 
$<$001$>$ corner display the low values of yield stress (Fig. \ref{Yield-stress}a). As the orientations move away from
the $<$001$>$ corner, an increase in compressive yield stress can be seen. The $<$111$>$ and $<$101$>$ oriented nanowires 
display the highest compressive yield stress values of 20.2 GPa and 13.1 GPa, respectively (Fig. \ref{Yield-stress}a). 
Contrary to compressive loading, the orientations close to $<$101$>$ corner display the low values of yield stress under 
tensile loading (Fig. \ref{Yield-stress}b). As the orientations move towards $<$111$>$ and $<$001$>$ corners, an increase 
in tensile yield stress can be seen in Fig. \ref{Yield-stress}b. Under tensile loading, the $<$111$>$ and $<$001$>$ 
oriented nanowires exhibits the highest values of yield stress. It is interesting to note that the $<$111$>$ oriented 
Cu nanowires display the highest value of yield stress under both tensile and compressive loading.

\section{Discussion}

\subsection{Orientation and loading mode dependent deformation mechanisms}

MD simulation results indicate that under compressive loading, the $<$100$>$, $<$103$>$ and $<$113$>$ oriented Cu 
nanowires deform by twinning mechanism, while the remaining orientations deform by dislocation slip. On the other hand 
all the nanowires deform by twinning mechanism under tensile loading. Full dislocation slip has not been observed in 
any of the orientations under tensile loading. The deformation by twinning in certain orientations has lead to the 
complete reorientation of the nanowires. These results shows that the $<$100$>$, $<$103$>$ and $<$113$>$ Cu nanowires 
deform by twinning mechanism, irrespective of the loading modes of tension and compression. Cu nanowires with remaining 
orientations exhibit tension-compression asymmetry in deformation mechanisms in terms of full dislocation slip under 
compressive loading and twinning under tensile loading. In the past, the tension-compression asymmetry in deformation 
mechanisms has been revealed in BCC nanowires \cite{Sainath-CMS16} and also in high symmetry $<$110$>$ and $<$111$>$ 
oriented FCC nanowires \cite{Park-JMPS,Lee-Nature}. However, the present study shows that the asymmetry in deformation 
mechanisms of FCC nanowires is not limited to $<$110$>$ or $<$111$>$ orientations, but it exists in many other orientations 
close to $<$110$>$ and $<$111$>$ corners of the standard stereographic triangle.

The orientation and mode of loading dependent deformation behaviour of FCC nanowires can be understood based on the 
Schmid factor ($m$) analysis \cite{Park-JMPS,Cai-Rev}. According to this analysis, if the Schmid factor of leading partial is 
higher than the trailing partial, then the deformation proceeds by the slip of partial dislocations or twinning \cite{Cai-Rev}. 
On the other hand, if the Schmid factor of trailing partials is higher than the leading partials, nucleation of trailing 
partial immediately follows the already nucleated leading partial and this results in deformation dominated by the slip 
of full or extended dislocations \cite{Cai-Rev}. The Schmid factor values for leading ($m_L$) and trailing ($m_T$) partials 
for some of the orientations of Cu nanowires along with the predicted and observed deformation mechanisms are presented 
in Table \ref{Schmid-factors}. It can be seen that for most of the orientations, the predicted mechanisms using the Schmid 
factor analysis are in good agreement with the observed mechanisms in the present study. The exception is being the 
$<$100$>$, $<$103$>$ and $<$113$>$ orientations under tensile loading. This exception can arise from the effects associated 
with the orientation of the side surfaces and also the nanowire size, which are not considered 
in the Schmid factor calculations \cite{Park-JMPS,Cai-Rev,Rohith-Philmag}. When the nanowire is enclosed by high energy 
side surfaces, it tries to minimize the surface energy by reorienting the surfaces, which is possible only when the 
deformation proceeds by twinning mechanism. Therefore in nanowire with high energy side surfaces generally twinning is 
preferred over dislocation slip, even though Schmid factor predicts otherwise. Similarly in small size nanowires partial 
slip/twinning is preferred, while increasing the size above certain value changes the deformation mode from twinning 
to full dislocation slip as reported in the previous study on $<$100$>$ Cu nanowires \cite{Rohith-Philmag,Cross-over}.

\begin{table}[h]
\caption{Schmid factor values for leading ($m_L$) and trailing ($m_T$) partial dislocations for some of the 
orientations of Cu nanowires along with the predicted and observed deformation mechanisms. The ratio of leading 
partial Schmid factor ($m_L$) to trailing partial Schmid factor ($m_T$) has been presented as $\alpha_M$.}
\label{Schmid-factors}
\centering
\scalebox{0.84}{
\begin{tabular}{| l | l | l | l | l | l | l | l} \hline
\rule{0pt}{3ex}
Nanowire & Loading & $m_L$ & $m_T$ & Predicted mechanism & Observed & $\alpha_M$ = \\
orientation & type &  &  &  & mechanism & $m_L$/$m_T$ \\ [3pt]
\hline
\rule{0pt}{3ex}
$<$100$>$ & Tension & 0.235 & 0.470 & Full slip & {\bf Twinning} & 0.5 \\
 & Compression & 0.470 & 0.235 & Twinning/partial slip & Twinning & 2.0\\ [3pt]
 \hline \rule{0pt}{3ex}
 $<$102$>$ & Tension & 0.42 & 0.42 & Full slip and Twinning & Twinning & 1.0 \\
 & Compression & 0.42 & 0.42 & Full slip and Twinning & Full slip & 1.0  \\ [3pt]
  \hline \rule{0pt}{3ex}
 $<$103$>$ & Tension & 0.38 & 0.47 & Full slip & {\bf Twinning} & 0.8 \\
 & Compression & 0.47 & 0.38 & Twinning & Twinning & 1.2  \\ [3pt]
 \hline \rule{0pt}{3ex} 
 $<$110$>$ \& & Tension & 0.470 & 0.235 &Twinning/partial slip & Twinning & 2.0 \\
 $<$212$>$ & Compression & 0.235 & 0.470 &Full slip & Full slip & 0.5  \\ [3pt]
 \hline \rule{0pt}{3ex}
$<$111$>$ & Tension & 0.31 & 0.155 & Twinning/partial slip & Twinning & 2.0 \\
 & Compression & 0.155 & 0.31 & Full slip & Full slip & 0.5  \\ [3pt]
  \hline \rule{0pt}{3ex}
 $<$112$>$ & Tension & 0.39 & 0.31  & Twinning/partial slip & Twinning & 1.3\\
 & Compression & 0.31 & 0.39  & Full slip & Full slip & 0.8\\ [3pt]
 \hline \rule{0pt}{3ex}
 $<$113$>$ & Tension & 0.39 & 0.43  & Full slip & {\bf Twinning} & 0.9\\
 & Compression & 0.43 & 0.39  & Twinning & Twinning & 1.1\\ [3pt]
 \hline \rule{0pt}{3ex}
 $<$214$>$ & Tension & 0.44 & 0.39  & Twinning/partial slip & Twinning & 1.1\\
 & Compression & 0.39 & 0.44  & Full slip & Full slip & 0.9\\ [3pt]
\hline 
\end{tabular} }
\end{table}

\begin{figure}[h]
\centering
\graphicspath{ {Figures/Chapter5/} }
\begin{subfigure}[b]{0.45\textwidth}
\includegraphics[width=\textwidth]{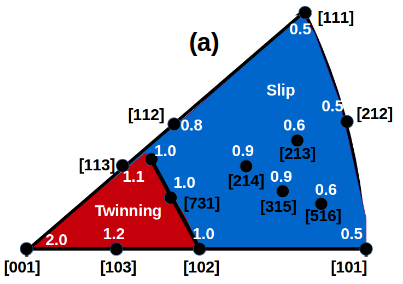}
\end{subfigure}
\begin{subfigure}[b]{0.45\textwidth}
\includegraphics[width=\textwidth]{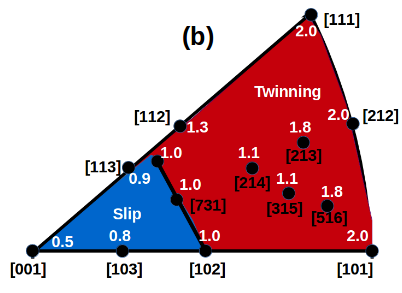}
\end{subfigure}
 \caption {The standard stereographic triangle showing two different regions, one with $\alpha_M > 1$ (red colour) 
and other with $\alpha_M < 1$ (light blue) under (a) compressive loading and (b) tensile loading. The black solid line 
between red and blue colours represent a boundary with $\alpha_M$ = 1. The parameter, $\alpha_M$ is defined as the 
ratio of Schmid factors of leading and trailing partial dislocations.} 
 \label{Triangle-m}
 \end{figure}

The above mentioned Schmid factor analysis can be generalized for all the orientations in standard stereographic triangle 
by defining a parameter $\alpha_M$, as the ratio of Schmid factors for leading and trailing partial dislocations (Table 
\ref{Schmid-factors}). The values of $\alpha_M$ for different orientations of Cu nanowires have been shown on standard 
stereographic 
triangle in Figs. \ref{Triangle-m}a and b under compressive and tensile loading, respectively. Based on the values of 
$\alpha_M$, the orientations in the triangle can be divided into two regions, one with $\alpha_M > 1 $ and the other with 
$\alpha_M < 1 $, separated by a boundary line with  $\alpha_M = 1 $. It can be seen that for the region where $\alpha_M > 1$, 
the Schmid factor for leading partial is higher than the trailing partials irrespective of the loading mode and as a result, 
the deformation by twinning/partial dislocation slip is favoured in orientations falling in this region. In contrast, for 
the region where $\alpha_M < 1$, the Schmid factor for trailing partial dislocations is higher than the leading partials. 
As a consequence, the deformation by full/extended dislocations is favoured in this region. For the orientations falling 
on the boundary with $\alpha_M = 1 $, both leading and trailing partials have equal Schmid factor values (e.g. $<$102$>$ 
orientation in Table \ref{Schmid-factors}) and the deformation can proceed either through twinning or extended dislocations 
or by the combination of these two \cite{TB-sliding-nature}. Therefore, if the nanowire orientation falls in the region with 
$\alpha_M > 1$, the deformation by twinning is preferred, while full dislocation slip is preferred if the orientation falls 
in the region with $\alpha_M < 1$. If it falls on the boundary line with $\alpha_M = 1$, both the mechanisms are equally 
probable \cite{TB-sliding-small}. The generalization of Schmid factor analysis based on the parameter $\alpha_M$ has an 
advantage as this method predicts the operative deformation mechanism in some arbitrary orientation simply based on its 
position on the triangle and loading mode.

\subsection{Tension-compression asymmetry in yield stress}

\begin{figure}[h]
\centering
\graphicspath{ {Figures/Chapter5/} }
\begin{subfigure}[b]{0.45\textwidth}
\includegraphics[width=\textwidth]{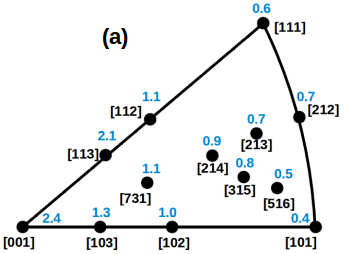}
\end{subfigure}
\begin{subfigure}[b]{0.45\textwidth}
\includegraphics[width=\textwidth]{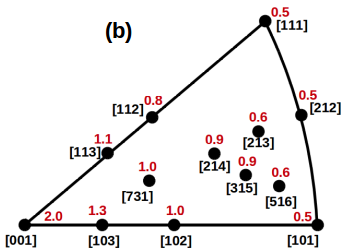}
\end{subfigure}
 \caption {The values of (a) yield stress asymmetry (b) Schmid factor asymmetry shown on the standard stereographic 
graphic triangle for different orientations of Cu nanowires. Yield stress asymmetry is defined as the ratio of yield stress 
in tension to that in compression (T/C), while the Schmid factor asymmetry is defined as the ratio of leading partial Schmid 
factor in compression to that in tension (C/T).} 
 \label{Asymmetry}
 \end{figure}

MD simulations results indicate that for most of the orientations, the values of yield stress were different under 
tensile and compressive loading i.e. the nanowires exhibit tension-compression asymmetry in yield stress. This asymmetry 
(r) can be better understood by considering the ratio of yield stress in tension to that in compression (T/C). Fig. 
\ref{Asymmetry}a shows the values of asymmetry factor 'r' for different orientations on the standard stereographic triangle. 
It can be seen that, the asymmetry values falls in the range 0.4-2.4 and for the orientations near $<$001$>$ corner, the 
asymmetry values are higher than the values for the orientations close to $<$110$>$ and $<$111$>$ corners. This suggest 
that, the orientations close to $<$001$>$ corner exhibits higher yield stress in tension than in compression, while the 
opposite behaviour (higher yield stress in compression than in tension) is observed in orientations close to $<$110$>$ and 
$<$111$>$ corners. Similar to the present study, the tension-compression asymmetry in yield strength has been observed in 
many FCC and BCC nanowires \cite{Lee-Nature,Park-JMPS,Sainath-CMS16,Weinberger-JMPS,Diao-NanoLett}. However, the present 
results shows that for the specific orientation of $<$102$>$, the values of yield stress under tensile and compressive 
loadings have been observed to be the same, thus indicating the absence of asymmetry. In order to understand this asymmetry 
in yield strength, different explanations have been given in literature \cite{Weinberger-JMPS}. Weinberger et al. 
\cite{Weinberger-JMPS} 
have pointed out that the presence of tensile surface stresses is responsible for the asymmetry in yield stress of FCC 
nanowires. The surface stress, which is generally tensile in nature, induces a compressive stress in the nanowires and 
makes them stronger in tension and weaker in compression \cite{Weinberger-JMPS}. However, the present results show that 
this explanation may not be valid for the orientations close to $<$110$>$ and $<$111$>$ corners. Further, the tension-
compression asymmetry has been reported in FCC bulk single crystals, where there is no presence of surface stresses 
\cite{Tschopp-APL,Salehinia-IJP,Xie-CMS-2014}. In the absence of surface stresses, it has been proposed that the stress 
normal to the slip direction and also the atomic stacking of the \{111\} slip planes, which is not symmetrical about the 
$<$110$>$ slip direction were responsible for the tension-compression asymmetry in FCC single crystals \cite{Tschopp-APL,
Salehinia-IJP,Xie-CMS-2014}. 

Different from the previous studies, the tension-compression asymmetry in FCC nanowires can be explained based on the 
Schmid factors of leading partial dislocations, which are different under tensile and compressive loading. Here, the 
Schmid factors of only leading partials is necessary because irrespective of the deformation mechanism (full slip/
partial slip/twinning), the first pop-in event during yielding is the nucleation of leading partial dislocation. 
Therefore, the yield stress is controlled mainly be the Schmid factor of leading partial dislocations. In order to 
relate the observed yield stress asymmetry (r) to Schmid factors of leading partials, the ratio of leading partial 
Schmid factor in compression to that in tension (C/T) has been taken as the Schmid factor asymmetry (s). Fig. 
\ref{Asymmetry}b shows the values of Schmid factor asymmetry (s) for different orientations on 
the standard stereographic triangle. It can be seen that for most of the orientations, the values of Schmid factor 
asymmetry (s) were approximately close to the values of yield stress asymmetry i.e. there is one to one correlation 
between yield stress asymmetry and Schmid factor asymmetry (Figs. \ref{Asymmetry}a and b). This 
correlation suggest that the asymmetry in yield stress of nanowires is arising mainly due to the different Schmid 
factors for leading partial dislocation under tensile and compressive loading. 

\section{Conclusions}

Molecular dynamics simulation results have shown that the deformation mechanisms in Cu nanowires vary significantly 
with crystallographic orientation and mode of loading. Under compressive loading, the orientations close to $<$100$>$ 
corner of a standard stereographic triangle i.e. $<$100$>$, $<$103$>$ and $<$113$>$ orientations deformed by twinning 
mechanism, while the remaining orientations deformed by full dislocation slip. On the other hand, all the orientations 
irrespective of their position on the stereographic triangle deformed by twinning mechanism under tensile loading. 
Further, the orientations close to $<$110$>$ and $<$111$>$ corners exhibited asymmetry in deformation mechanisms in 
terms of full dislocation slip under compressive loading and twinning under tensile loading. Irrespective of twinning 
or full dislocations, the orientations falling in the interior of the standard stereographic triangle deformed on a 
single slip/twin system, while the orientations falling on the border deformed on multiple slip/twin systems. In 
addition to deformation mechanisms, it has been observed that the Cu nanowires display tension-compression asymmetry 
in yield stress. The orientations close to $<$001$>$ corner exhibits higher yield stress in tension than in compression, 
while higher 
yield stress in compression than in tension has been observed in orientations close to $<$110$>$ and $<$111$>$ corners. 
Interestingly, for the specific orientation of $<$102$>$, the values of yield stress under tensile and compressive 
loading were found to be same, thus indicating the absence of yield stress asymmetry in $<$102$>$ Cu nanowire. 
Irrespective of loading mode, the $<$111$>$ oriented Cu nanowires displayed the highest yield stress among all the 
orientations investigated. The tension-compression asymmetry in deformation mechanisms has been explained 
based on the parameter $\alpha_M$, defined as the ratio of Schmid factors for leading and trailing partial dislocations.
For the nanowire orientations falling in the region with $\alpha_M > 1$, the deformation by twinning is preferred, 
while full dislocation slip is observed for the orientations falling in the region with $\alpha_M < 1$. For the 
orientations falling on the boundary line with $\alpha_M = 1$, both the mechanisms are equally probable. However, 
the exception is being the $<$100$>$, $<$103$>$ and $<$113$>$ orientations under tensile loading, where twinning 
has been observed under tensile and compressive loading. Similarly, the asymmetry in yield values has been attributed 
to the different Schmid factor values for leading partial dislocations under tensile and compressive loading.


\end{document}